\documentclass[aps,prd,onecolumn,amssymb,eqsecnum,nofootinbib]{revtex4}

\usepackage{amsmath}
\usepackage{amssymb}

\newcommand{\bs}[1]{\boldsymbol{#1}}

\usepackage{epsf}  
\usepackage{graphicx}

\begin{document}

\title{NP Stokes fields for radio astronomy}
\author{Ezra T.~Newman}
\email{newman@pitt.edu}
\affiliation{
Department of Physics and Astronomy, University of Pittsburgh, Pittsburgh,
PA 15260}

\author{Richard H.~Price}
\email{Richard.Price@utb.edu}
\affiliation{Center for Gravitational Wave Astronomy, 
80 Fort Brown, Brownsville, TX. 78520}

\begin{abstract}
  The spin weighted spherical harmonic (SWSH) description of angular
  functions typically is associated with the Newman-Penrose (NP) null
  tetrad formalism.  Recently the SWSH description, but not the NP
  formalism, has been used in the study of the polarization
  anisotropy of the cosmic microwave background. Here we relate this
  application of SWSHs to a description of electromagnetic radiation
  and polarization in the NP formalism.  In particular we introduce NP
  Stokes fields that are the NP equivalent of the Stokes parameters.
  In addition to giving a more coherent foundation for the recent
  cosmological SWSH application, the NP formalism aids in the
  computation of the Lorentz transformation properties of polarization.
\end{abstract}

\maketitle

\section{Introduction}\label{sec:intro}

To describe relativistic fields, the null tetrad, or
Newman-Penrose(NP) formalism\citep{NP}, uses a set of four null
spacetime vectors ${\vec\ell},\vec{n}, \vec{m}, \vec{m^*}$ defined at
every point in some region of spacetime. In the NP formalism, rather
than deal with the components of a tensor, say the Maxwell tensor
$F_{\mu\nu}$ for electromagnetism, one uses the projections of tensors
on this tetrad. Because the tetrad field is chosen to satisfy certain
properties, it turns out that the NP fields, the projected quantities,
are convenient for the mathematics of radiation fields.  Since the
tetrad legs themselves have angular properties, projection with them
adds some angular dependence not inherent in the physical field
themselves. This additional angular dependence means that in general
NP fields should not be expanded in ordinary spherical harmonics.
Rather, all the simplicity of spherical harmonic expansions is
regained if NP fields are expanded in sets of angular functions called
spin weighted spherical harmonics (SWSHs)\cite{Goldbergetal67}.

SWSHs have recently played a significant role in research on
cosmological anisotropies. 
The directional nature of linear polarization complicated comparisons
of linear polarization in different sky directions, and limited polarization
studies to small angular comparisons.
In 1996 Seljak\cite{Seljak96} found that certain combinations of the
Stokes parameters $Q$ and $U$ were particularly well suited to dealing
with the angular properties of linear polarization.
This insight was still limited, as had been earlier work, to
comparisons over small angular regions. That restriction was removed
soon after in the breakthrough work by Seljak and
Zaldarriaga\cite{SeljaZaldarriaga97PRL,ZaldSeljak97} in which SWSHs
were introduced. There soon followed papers\cite{HuWhite97,NgLiu} with
further mathematical details of the use of SWSHs for cosmological
anisotropies.

This application of SWSHs to cosmology exploited their connection to
the rotation group \cite{GelfanMinlosShapiro58}, rather than the role
that the SWSHs typically play in physics.  To our knowledge, none of
the papers on the application of SWSHs to cosmological anisotropies
makes reference to the NP formalism itself. One of our purposes here
is to show the very natural connection that exists.
In this paper we
introduce a NP formalism for linear polarization and for Stokes
parameters, and we show  in this formalism how SWSHs very naturally
arise. We argue, furthermore, that the NP formalism has advantages
beyond that of giving an explanatory view of a technique already in
use.

Since we will only be considering a description of the fields at an
observer location, not the propagation of the field, it is
appropriate, for simplicity, to limit ourselves to the notation of
special, not general relativity.  But we make this simplification with
some regrets since the NP formalism is extremely useful for dealing
with the propagation of fields in curved spacetime. For some problems
in curved spacetime, in fact, it is almost indispensable.

We  will use the standard special relativistic metric 
with sign convention +\,-\,-\,- and with the speed of light $c$ set to unity, so
that 
in Minkowski coordinates
\begin{equation}
  \label{eq:srtmetric}
  ds^2=dt^2-dx^2-dy^2-dz^2\,.
\end{equation}
Greek letters  $\{\mu,\nu...\}$ will indicate
indices on 4-vectors, and Latin letters 
$\{i,j,...\}$
will indicate  
indices on 3-vectors, also called ``spatial vectors.''
Arrows over symbols indicate 4-vectors; boldface symbols will be used for 
3-vectors.
For 3-vectors, such as the electric
and magnetic vectors, subscript and superscript indices are equivalent since
the bases used are orthonormal and the 3-metric is $\delta_{ij}$. The 3-vector 
components are equivalent to the spatial contravariant (superscript) components of 
a corresponding 4-vector. We use the same root symbol, $\ell$, for the multipole
index and for one of the null legs of the NP tetrad. The difference will be clear
from context, since in the latter usage it will either appear as a 4-vector $\vec{\ell}$
or will be represented by its components $\ell^\mu$.
\begin{figure}[h]
  \centering
\includegraphics[width=.3\textwidth]{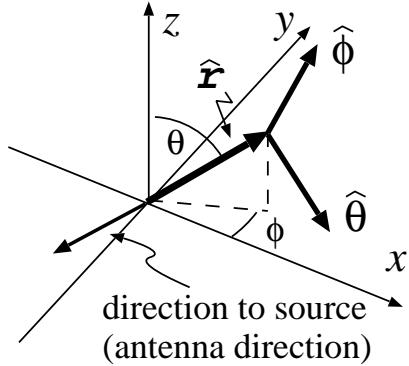}   
  \caption{Direction of wave propagation and basis vectors.}
  \label{fig:pvector}
\end{figure}


\section{The NP formalism}
\label{sec:NPform}
The situation we consider is that of an observer and an antenna at our coordinate
origin. Radiation from all directions arrives at the 
origin, although not necessarily isotropically. 
This radiation can be considered to be a distribution of plane waves, and
we  restrict attention 
to  waves
propagating in a narrow range of directions around some particular
direction $\widehat{\bs r}$. As shown in Fig.~\ref{fig:pvector}, this
direction can also be specified with the angles $\theta$ and $\phi$.
The Cartesian components of the vector $\widehat{\bf r}$
can be written as
\begin{equation}
  \{\widehat{r}_x,\widehat{r}_y,\widehat{r}_z\}=\{\sin\theta\cos\phi,\sin\theta\sin\phi,
\cos\theta\}\,.
\end{equation}
It will be very useful to have, in addition to the Cartesian basis, a
basis consisting of $\widehat{\bs r}$ and the two orthogonal unit
vectors tangent to the celestial sphere, $\widehat{\bs\theta}$ and
$\widehat{\bs\phi}$, as shown in Fig.~\ref{fig:pvector}.

The direction in which an antenna must be pointed to receive this 
radiation is, of course, 
$-\widehat{\bs r}$, so that the sky position the observer would assign
to the source of this radiation is the antenna direction $\theta_{(A)},\phi_{(A)}$
given by $\theta_{(A)}\equiv\pi-\theta$ and
$\phi_{(A)}\equiv\phi+\pi$.

We now use the null tetrad to help in the discussion of radio waves propagating 
in the $\widehat{\bs r}$ direction.
The NP null tetrad consists of two real null 4-vectors
$\vec{\ell}$ and $\vec{n}$, and the complex null vector $\vec{m}$,
which can usefully be considered to be equivalent to two null
4-vectors $\vec{m}$, and its complex conjugate $\vec{m}^*$.  The four
null vectors must have all dot products to be zero except for the dot
products in the following normalization conditions
\begin{equation}
  \label{eq:normcond}
  \vec{\ell}\cdot\vec{n}=1\quad\quad   \vec{m}\cdot\vec{m}^*=-1\ .
\end{equation}

In terms of the notation just defined, we choose our NP tetrad
to be the null vectors with the following contravariant components
\begin{eqnarray}
  \ell^{\mu}=\{\ell^0,\ell^j\}&=&2^{-1/2}
\{1,\widehat{r}_j\}\label{tetraddef1}\\
  n^{\mu}&=&2^{-1/2}
\{1,-\widehat{r}_j\}\label{tetraddef2}\\
  m^{\mu}&=&2^{-1/2}
\{0,-\widehat{\theta}_j-i\widehat{\phi}_j\}\,.\label{tetraddef3}
\end{eqnarray}
The Minkowski components of these tetrad legs are
\begin{eqnarray}
\ell^{\mu}&=&\{\ell^0,\ell^x,\ell^y,\ell^z\}=2^{-1/2}\label{minkcomp1}
\{1,\;
\sin{\theta}\cos{\phi},\;\sin{\theta}\sin{\phi},\;\cos{\theta}\}\label{minkcomp2}\\
  n^{\mu}&=&2^{-1/2}\{1,\;
-\sin{\theta}\cos{\phi},\;-\sin{\theta}\sin{\phi},\;-\cos{\theta}
\}\\
  m^{\mu}&=&2^{-1/2}\{0,\;
-\cos{\theta}\cos{\phi}+i\sin{\phi},\;
-\cos{\theta}\sin{\phi}-i\cos{\phi},\;\sin{\theta}\}\,.\label{minkcomp3}
\end{eqnarray}
It should be noted that all physical fields in the beam can be
functions of spacetime location $x^\mu$ only through the combination
$k_\mu x^\mu\propto t-r_jx^j$, where $\vec{k}$ is the propagation
4-vector of the beam. It follows that for the beam propagating in the
$\widehat{\bs r}$ spatial direction $k_\mu\propto\{1,-r_j\}$, and
$k^\mu\propto\{1,r_j\}$. Thus our tetrad leg $\vec{\ell}$ is
proportional to the propagation 4-vector $\vec{k}$.

 In the NP formalism the six independent pieces of information about
 the electromagnetic field at a spacetime point are represented by the following
complex projections\cite{Fmunuconvention} of the Maxwell tensor $F_{\mu\nu}$
 (see the Appendix of Ref.~\citep{NP}):
\begin{eqnarray}
  \Phi_0&\equiv& F_{\mu\nu}\ell^\mu m^\nu= \textstyle{\frac{1}{2}}\left[   
-E^{{{\hat{\theta}}}}+B^{{{\hat{\phi}}}}-i\left(E^{{{\hat{\phi}}}}+B^{{{\hat{\theta}}}}\right)
\right]\nonumber\\
\Phi_1&\equiv &\textstyle{\frac{1}{2}}F_{\mu\nu}\left(
\ell^\mu n^\nu-m^\mu m^{*\nu}\right)=
-\textstyle{\frac{1}{2}}
\left[E^{\hat{r}}+iB^{\hat{r}} \right] \label{eq:Phidefs}\\
  \Phi_2&\equiv& F_{\mu\nu}m^{*\mu} n^\nu=\textstyle{\frac{1}{2}}
\left[   E^{{{\hat{\theta}}}}+B^{{{\hat{\phi}}}}
+i\left(-E^{{{\hat{\phi}}}}+B^{{{\hat{\theta}}}} \right)\right]\,.\nonumber
\end{eqnarray}
Here the components $E^{\hat{r}}, B^{\hat{\theta}}$ etc., represent the ordinary components of
the electric and magnetic fields with respect to the orthonormal basis system 
shown in Fig.~\ref{fig:pvector}.

Any of the  NP projections is characterized by two important properties. The
first is its 
{\em spin weight}, the number of $\vec{m}$ vectors
used in the projection, minus the number of $\vec{m}^*$s used. 
This spin weight is connected to the spin weight of the SWSHs.  In
fact, SWSHs consist of sets of spherical harmonics of specific
spin weights.  A spin weight $s$ quantity must be expanded in
spin weight $s$ spherical harmonics to realize the simplicity
mentioned in Sec.~\ref{sec:intro}. In view of the close connection
of projected fields of specific spin weight, and SWSHs, it is understandable
that the use of SWSHs in physics is
traditionally  associated with the NP formalism. 
The properties of SWSHs have been described 
in detail both in connection with the NP formalism\cite{Goldbergetal67,NewmanPenrose66}
and for use with cosmological anisotropies\cite{ZaldSeljak97,HuWhite97,NgLiu}.
We will not need those details here; we need only the motivation that 
the SWSHs have turned out to be important in the description of the linear
polarization of the CMB. 

The second important property of NP fields is that each has a definite
{\em boost weight} (also called {\em conformal weight}). This is the
number of $\vec{\ell}$\,s minus the number of $\vec{n}$\,s that are
used in projecting the NP field. Just as spin weight determines
rotation properties, the boost weight determines properties under
Lorentz transformations.  By using the NP description of linear
polarization, as we show in Sec.~\ref{sec:Lorentz}, it is straightforward 
to calculate the changes measured under a general Lorentz boost.

\section{Stokes parameters in the NP formalism} 
\label{sec:NPrad}

We note that  the plane waves traveling in the direction
shown in Fig.~\ref{fig:pvector} must have their  
Poynting vector directed in the $\widehat{\bs r}$ direction, and hence must
have $E^{\hat{r}}=B^{\hat{r}}=0$,
$E^{\hat{\theta}}=B^{\hat{\phi}}$ and $E^{\hat{\phi}}=-B^{\hat{\theta}}$.
In this case, $\Phi_0$ and $\Phi_1$ vanish, and all the electromagnetic
information is carried in $\Phi_2$ which reduces to
\begin{equation}
  \label{eq:Phi0rad}
\Phi_2=
E^{\hat{\theta}}-iE^{\hat{\phi}}\,.  
\end{equation}

In the description of Stokes parameters a complex representation 
of the electric field is typically used. Because the NP quantities 
have their own complex nature, we avoid the more typical approach and 
describe a frequency component of  the electric field as
\begin{equation}\label{calEjdef}
  E^{\hat{\theta}}={\cal E}^\theta \cos{(\omega t+\delta_\theta(\omega))}\quad\quad\quad
  E^{\hat{\phi}}={\cal E}^\phi \cos{(\omega t+\delta_\phi(\omega))}\,.
\end{equation}
We next define  a field that is  phase shifted by $\pi/2$:
\begin{multline}\label{phaseshift}
\widetilde{\Phi_2}= 
  {\cal E}^\theta \cos{(\omega t+\delta_\theta(\omega)+\pi/2)}
-i {\cal E}^\phi \cos{(\omega t+\delta_\phi(\omega)+\pi/2)}
\\
= -
  {\cal E}^\theta \sin{(\omega t+\delta_\theta(\omega))}
+i {\cal E}^\phi \sin{(\omega t+\delta_\phi(\omega))}\,.
\end{multline}

To construct the NP quantities for radio astronomy we average over the range of 
frequencies of interest to form the following ``NP Stokes fields''
\begin{equation}\label{allcals}
  {\cal I}=2\Bigl\langle\Phi_2\Phi_2^*
\Bigr\rangle
\quad\quad\quad
  {\cal S}=2\Bigl\langle\Phi_2\Phi_2
\Bigr\rangle
\quad\quad\quad
  {\cal V}=2i\Bigl\langle{\Phi_2}\widetilde{\Phi_2}^*
\Bigr\rangle\,.
\end{equation}
The angle brackets $\langle\rangle$ denote the average over some 
frequency interval. Unless the radiation of interest is totally unpolarized
we assume that $\delta_\theta\equiv\langle\delta_\theta(\omega)\rangle$
and  $\delta_\phi\equiv\langle\delta_\phi(\omega)\rangle$ are not both zero.

The first of these quantities is straightforward to evaluate with 
Eq.~\eqref{eq:Phi0rad} to show that 
\begin{equation}
  {\cal I}=2\left\langle(E^{\hat{\theta}})^2
+(E^{\hat{\phi}})^2
\right\rangle =({\cal E}^\theta)^2+({\cal E}^\phi)^2\equiv I.
\end{equation}
The last symbol in this equation is the usual Stokes parameter
$I$. (See, e.g., Sec.~7.2 of Ref.~\cite{Jackson3rd}).  Up to
multiplicative factors this quantity is clearly just the intensity of
radiative energy, i.e., the magnitude of the Poynting vector. The NP
quantity ${\cal I}$ has spin weight zero, in complete accord with what
one would expect on physical grounds, since the intensity, or
temperature, is a scalar under rotations.

In terms of the electric field, the NP quantity ${\cal S}$ has the
form
 \begin{equation}
  {\cal S}=2\left\langle(E^{\hat{\theta}})^2
-(E^{\hat{\phi}})^2
-2i E^{\hat{\theta}}E^{\hat{\phi}}
\right\rangle    =   
({\cal E}^\theta)^2-({\cal E}^\phi)^2-
2i {\cal E}^\theta {\cal E}^\phi \cos{(\delta_\theta - \delta_\phi )}
 =Q-iU\,,
\end{equation}
where $Q$ and $U$ are the usual Stokes parameters of the radiation. (See,
e.g., Sec.~7.2 of Ref.~\cite{Jackson3rd}). 
Unlike ${\cal I}$, the ${\cal S}$ field has spin weight -2, which means that
an expansion of ${\cal S}$ in functions on the sphere is best carried out
with spin weight -2 harmonics\cite{orplus2}. The fact that this 
spin weight -2 NP Stokes field
(or its spin weight +2 complex conjugate) carries the information about
linear polarization explains the 
applicability of spin weight $\pm2$ SWSHs to the studies of the linear 
polarization of the CMB.

In terms of the electric field the third of our NP Stokes fields is
\begin{equation}\label{calVdef}
  {\cal V}=2{\cal E}^\theta {\cal E}^\phi \sin{(\delta_\phi - \delta_\theta )}= V\,,
\end{equation}
where $V$ is the last of the usual Stokes parameters.  Like ${\cal I}$, and
unlike ${\cal S}$, the field ${\cal V}$ has spin weight 0. The
mathematical reason for this is that the Stokes parameter $V$ is
invariant with respect to rotations about the direction of
propagation. Like ${\cal I}$, the circular polarization measure 
 ${\cal V}$  should be expanded in ordinary spherical harmonics.
(We mention, only as an aside, that at present cosmological models do not
include processes that produce circular polarization.)

 All three NP Stokes fields have boost weight -2. We will see below
 that the boost weight of an NP quantity indicates the quantity's
 transformation property under a Lorentz boost. In particular a
 boost weight $w$ quantity is multiplied by the Doppler shift (as
 defined in Sec.~\ref{sec:Lorentz}) raised to the power $w$.  Thus,
 all the Stokes fields must be multiplied by the square of the Doppler
 shift under a boost. This is in accord with the fact that all the
 Stokes fields are quadratic in the transverse electric fields, which
 themselves are multiplied by a Doppler shift factor under a boost.
 (An alternate viewpoint is that every photon is Doppler shifted, and
 the rate of photon arrival is Doppler shifted, hence all radiative
 intensities gets doubly Doppler shifted.)

\section{Lorentz tranformations} 
\label{sec:Lorentz}

Here we consider the behavior of the NP radiative quantities under a
Lorentz transformation, the transformation of measured quantities
associated with a change of observer rest frame, in particular for a
``boost,'' a transformation from the reference frame of an observer to
the frame of a relatively moving observer. 
Transformations of Stokes parameters have been considered by Challinor
and van Leeuwen \cite{ChallinorvanLeeuwen}, who did not use SWSHs
or the NP formalism; rather they confined their study to small angular
scales. We show here that the NP formalism removes the restriction of
small angular scales for Lorentz transformations, just as it does for
angular correlations.

The complete set of Lorentz transformations of reference frame can 
be considered to be equivalent to the transformations of the null tetrad 
that maintain the conditions in Eq.~\eqref{eq:normcond}. 
The group of transformations of the tetrad is usefully divided into
several separate
subclasses\cite{Kinnersley1969,Price1972II}, as follows.\\
Class I:
\begin{eqnarray}
  \vec{n}^{\,\prime}&= & \vec{n}  \label{eq:I1}\\
\vec{m}^{\,\prime}&=&\vec{m}+b\,\vec{n}\label{eq:I2}\\
  \vec{\ell}^{\,\prime}&= & \vec{\ell}+b\,\vec{m}^*+b^*\vec{m}+bb^*\vec{n}\,,\label{eq:I3}
\end{eqnarray}
Class II:
\begin{eqnarray}
  \vec{\ell}^{\,\prime}&=&\vec{\ell}\label{firstclass1}\\
  \vec{m}^{\,\prime}&=&\vec{m}+a\vec{\ell}\label{firstclass2}\\
  \vec{n}^{\,\prime}&=&\vec{n}+a\,\vec{m}^*+{a}^*\vec{m} +a{a}^*\vec{\ell}
\label{firstclass4}\,,
\end{eqnarray}
Class III:
\begin{equation}\label{eq:III}
  \vec{m}^\prime=  e^{i\lambda}\vec{m}\,,\quad\quad\vec{\ell}^{\,\prime}=\vec{\ell}\,,
\quad\quad\vec{n}^{\,\prime}=\vec{n}\,,
\end{equation}
Class IV:
\begin{equation}
  \vec{m}^\prime=  \vec{m}\,,\quad\quad\vec{\ell}^{\,\prime}=K\vec{\ell}\,,
\quad\quad\vec{n}^{\,\prime}=K^{-1}\vec{n}\,.\label{eq:IV}
\end{equation}
The two complex parameters $a,b$ and the two real parameters $\lambda$
and $K$ contain the six degrees of freedom of the Lorentz group. They
can  be related to the usual boost and rotation
parameters of the coordinate basis vectors of observers' reference frames if a
specific association is made between the null tetrad $\{\vec{\ell},
\vec{n},\vec{m},\vec{m}^*\}$ and the reference frame of an observer.

Since the radiative quantities defined in Eqs.~\eqref{eq:Phidefs}
are constructed from the tetrad legs, the transformations in Eqs.~\eqref{eq:I1}--
\eqref{eq:IV} induce transformations to new radiative quantities $\Phi_0^{\prime},
\Phi_1^{\prime},\Phi_2^{\prime}$, as follows.\\
Class I:
\begin{eqnarray}
 \Phi_0^\prime&=&\Phi_0+2b\Phi_1+b^2\Phi_2\label{ClassIPhifirst}\\
 \Phi_1^\prime&=&\Phi_1+b\Phi_2\\
 \Phi_2^\prime&=&\Phi_2
\end{eqnarray}
Class II: 
\begin{eqnarray}
  \Phi_0^\prime&=&  \Phi_0\\
  \Phi_1^\prime&=&\Phi_1+a^*\Phi_0\\
  \Phi_2^\prime&=&\Phi_2+2a^*\Phi_1+a^{*2}\Phi_0
\end{eqnarray}
Class III:
\begin{equation}
  \Phi_0^\prime=e^{i\lambda}\Phi_0\,, \quad\quad
  \Phi_1^\prime=\Phi_1\,, \quad\quad
  \Phi_2^\prime=e^{-i\lambda}\Phi_2 
\end{equation}
Class IV:
\begin{equation}
  \Phi_0^\prime=K\Phi_0\,, \quad\quad
  \Phi_1^\prime=\Phi_1\,, \quad\quad
  \Phi_2^\prime=K^{-1}\Phi_2 \label{ClassIPhilast}\,.
\end{equation}

We are now in a position to apply this mathematical infrastructure to
the radiation problem. We consider two relatively moving observers
${\cal O}$ and ${\cal O}^{\,\prime}$,  with 4-velocities $\vec{u}$ and
$\vec{u}^{\,\prime}$, and  we focus our attention on a particular beam of radiation.

The interpretation of the NP Stokes fields given in
Secs.~\ref{sec:NPrad} requires that $\vec{\ell}$\,
be in the null direction of propagation of the radiation and that
$\vec{\ell}\cdot\vec{u}=1/\sqrt{2\;}$.  Since we want both observers
to be making the same physical measurements, only in different frames,
we must have that $\vec{\ell}^{\,\prime}$ is also in the null
direction of propagation 
(the same spacetime direction as 
 $\vec{\ell}$\;)
and that
$\vec{\ell}^{\,\prime}\cdot\vec{u}^{\,\prime}=1/\sqrt{2\;}$.  From the
requirement that  $\vec{\ell}$ and $\vec{\ell}^{\,\prime}$ be in the
same null direction, we infer that the tetrads of the two observers
must be related only by transformations of Classes II, III, and IV,
since the transformations of Class I in
Eqs.~\eqref{eq:I1}--\eqref{eq:I3}, change the direction of
$\vec{\ell}$. Equivalently, $b$ must vanish and the relationship of
the measurements of the two observers must depend only on the real
parameters $K,\lambda$ and on the complex parameter $a$.

An interesting conclusion now follows.   As pointed out at
the start of Sec.~\ref{sec:NPrad}, $\Phi_0=\Phi_1=0$. The  
transformations in
Eqs.~\eqref{ClassIPhifirst}--\eqref{ClassIPhilast},
with $b=0$, then tell us that
$\Phi_0^\prime=\Phi_1^\prime=0$. This, of course, must be the case on
physical grounds. Furthermore, these relations tell us that the most
general nontrivial transformation of $\Phi_2$ is
$\Phi_2^\prime=e^{-i\lambda}K^{-1}\Phi_2$. For the NP Stokes fields
this means that
\begin{equation}\label{NPStokesxform}
  {\cal I}^\prime=K^{-2}{\cal I}\,,
\quad\quad
  {\cal S}^\prime=e^{-2i\lambda}K^{-2}{\cal S}\,,
\quad\quad
  {\cal V}^\prime=K^{-2}{\cal V}\,.
\end{equation}

In a Lorentz transformation, then, the polarization properties in
${\cal S}$ are affected by both $K$ and by the  $\lambda$ parameter of a Class III
transformation, while the other two NP Stokes fields, as well as the
magnitude of ${\cal S}$, are affected only by $K$. 
The detailed relationship of the parameters $K$ and $\lambda$, and 
the frames of observers, are given 
in Appendix~\ref{genLXform}. Here we only give the result for $K$.
If
${\bs\beta}$ is the 3-velocity of a frame  ${\cal O}^\prime$, as
observed by ${\cal O}$, then 
\begin{equation}
  K
=\frac{\sqrt{1-\beta^2\;}}{1-{\bf\hat{r}}\cdot{\bs\beta}}\,.
\end{equation}
This is the  well known Doppler factor that relates observations of radiation by
the two observers, and is the  ratio $E/E^\prime$ of the observed
energies.

Although the formal transformation properties in
Eq.~\eqref{NPStokesxform} are correct, they cannot directly be related
to observations of the CMB. The Stokes parameters, and the NP Stokes
fields, measure radiative power per unit area. Radio telescopes,
however, measure radiative power per unit area {\em per unit frequency
  interval $d\nu$, per unit solid angle $d\Omega$}. In considering
transformation properties we must take into account that
\begin{equation}\label{nuOmegaxform}
  \nu^\prime=K^{-1}\,\nu\quad\quad  d\Omega^\prime=K^2\,d\Omega\,.
\end{equation}
Thus, a radio telescope does not measure e.g.\,, the Stokes parameter
$I$, but rather the specific intensity 
\begin{equation}
  I_\nu\equiv \frac{dI}{d\nu d\Omega}\,.
\end{equation}
From the transformation properties in Eqs.~\eqref{NPStokesxform} and
\eqref{nuOmegaxform} it follows that the specific intensity $I_\nu$
transforms according to 
\begin{equation}
 I_\nu^\prime=K^{-3}I_\nu\,. 
\end{equation}
We can apply this
to the NP Stokes fields and can introduce measureable quantities ${\cal I}_\nu$,
${\cal S}_\nu$, ${\cal V}_\nu$, defined per unit frequency per unit
solid angle. The transformation laws for these quantities are those of
Eq.~\eqref{NPStokesxform}, with $K^{-2}$ replaced by $K^{-3}$. 

These
specific intensities are directly related to what is measured by
radiotelescopes, but it is simple, and practical, to construct other
quantities directly related to telescope observables. The specific 
intensities can be combined with the frequency at which the specific 
intensity is measured to form quantities  $I_\nu/\nu^N$, and the 
equivalent for NP Stokes fields. For these fields the transformation
properties are
\begin{equation}\label{genNPStokesxform}
 \left({\cal I}_\nu/\nu^N\right)^\prime=K^{N-3}\left({\cal I}_\nu/\nu^N\right)
\quad\quad
 \left({\cal S}_\nu/\nu^N\right)^\prime=e^{-2i\lambda}K^{N-3}\left({\cal S}_\nu/\nu^N\right)
\quad\quad
 \left({\cal V}_\nu/\nu^N\right)^\prime=K^{N-3}\left({\cal V}_\nu/\nu^N\right)\,.
\end{equation}

In cosmology, the Lorentz transformation properties of radiation
quantities are important for transforming observations to a ``rest
frame'' of the CMB. In principle this is the frame that must be used
for comparisons of observations and theories of early universe
processes. The general prescription for finding this frame is to identify
a dipole in the radiation. A transformation can then be made to a frame
in which this dipole vanishes. The process, however, is not unique. One
can, for example, choose to elminate the dipole in circular polarization
rather than intensity. 

The problem of finding a rest frame is essentially the problem of
understanding how the radiation gives a representation of the Lorentz
group. The general theory of such representations can be based on the
work of Gel'fand, Graev and Vilenkin\cite{GGV} and is sketched in
Appendix \ref{app:GGV}. With this theory, Lorentzian vectors and
tensors are constructed from the multipoles, the 2-sphere integrals of
radiation quantities.  
This issue of a choice of rest frames is taken up in greater detail in 
Appendix~\ref{app:GGV}.


\section{Summary and Conclusions} 
\label{sec:Conc}

The introduction\cite{ZaldSeljak97} of SWSHs to the analysis of
polarization anisotropy followed from a consideration of the explicit
rotation properties of the Stokes parameters $Q,U$ and to the
subsequent realization that a combination of the two had convenient
transformation properties under rotation. We have shown here that the
NP formalism leads very naturally to an NP formulation of the Stokes
parameters, and that in this formulation the use of spin weight $\pm2$
SWSHs is immediate. We have also shown the convenience of the NP
formalism for ``boosts,'' the transformtion of radiation properties
from an oberver's frame to the frame of a relatively moving observer.
More specifically, we have shown that the NP
formalism allows for the computation of boosts of any radiation quantity
without the limitation of small angular scale.

The convenience of the NP formalism is not a coincidence. Its
mathematical underpinnings are deeply rooted in the structure of
spacetime, so the formalism is very naturally suited to the
description of fields that propagate at the speed of light. This can
be taken as the reason that Stokes parameters arise simply as fields
quadratic in the NP field describing electromagnetic radiation. This
even more certainly explains the relative simplicity of boost
transformations. The brief discussion given of the nonuniqueness of
the ``rest frame'' of the CMB is closely associated with the
convenience of the NP formalism for boosts.  

Since the advantages, or conveniences, follow from the appropriateness
of the NP formalism for fields propagating along null directions in
spacetime, they apply also to non-electromagnetic fields. In
particular, gravitational waves, whether the standard
transverse-traceless waves of Einstein's theory, or the full set of
six possible polarization states, all have gravitational Stokes
parameters that follow from fields quadratic in the NP projections of the 
Riemann tensor.

Lastly, we mention that the NP formalism is very naturally suited to
calculations of the propagation of radiation in curved spacetime.  In
the cosmological context this means that the NP formalism should
greatly simplify, for example, the calculation of the effect on
polarization of gravitational lensing or inhomogeneous cosmological
expansion.

\section{Acknowledgment}
One of us (RHP) gratefully acknowledges support for this work under
NSF grants PHY-0554367 and by the Center for Gravitational Wave
Astronomy. We thank Arthur Kosowsky for very helpful discussions about
the cosmological background for this work, and Teviet Creighton and Fredrick
Jenet for discussions of statistical issues in connection with anisotropies.

\appendix

\section{The general Lorentz transformation}\label{genLXform}

To relate the value of $K$ to the usual parameterization of a Lorentz
transformation we consider, as in Sec.~\ref{sec:Lorentz}, a frame ${\cal O}^\prime$
observed in frame ${\cal O}$ to be moving with 3-velocity ${\bs\beta}$.
The  propagation 4-vector $\vec{k}$ of a
plane electromagnetic wave (equivalently, the 4-momentum of a photon)
 must be parallel to $\vec{\ell}$, so we can write
$\vec{k}=\kappa\,\vec{\ell}$, where $\kappa$ is proportional to the
energy $E$ of a ``photon'' in frame ${\cal O}$.  
With $\vec{\ell}^{\,\prime}=K\vec{\ell}$, we have from this
\begin{equation}
  \frac{E^{\,\prime}}{E}=\frac{\vec{k}\cdot\vec{u}^{\,\prime}}{\vec{k}\cdot\vec{u}}
=\frac{\vec{\ell}\cdot\vec{u}^{\,\prime}}{\vec{\ell}\cdot\vec{u}}
=K^{-1}\,\frac{\vec{\ell}^{\,\prime}\cdot\vec{u}^{\,\prime}}{\vec{\ell}\cdot\vec{u}}
=K^{-1}\,.
\end{equation}

To get the specific dependence of $K$ on the relative direction of motion, and
the direction of observation, we can use a straightforward evaluation
of $\vec{\ell}\cdot\vec{u}^\prime$ in the ${\cal O}$ frame, in which 
$\vec{\ell}$ has the contravariant components in Eq.~\eqref{tetraddef1}, 
and $\vec{u}^{\,\prime}$
has the contravariant components
\begin{equation}
  u^{\prime\mu}=\gamma\{1,\beta^j\}\,,
\end{equation}
 where $\gamma$ is the usual
Lorentz factor $1/\sqrt{1-\beta^2\;}$. From these components we get
\begin{equation}\label{eq:elldotu}
  \vec{\ell}\cdot\vec{u}^{\,\prime}=\gamma
\left(1-{\bf \hat{r}}\cdot{\bs\beta}\right)/\sqrt{2\;}.
\end{equation}
But we have  $\vec{\ell}= K^{-1}\vec{\ell}^{\,\prime}$ so that $\vec{\ell}\cdot
\vec{u}^{\,\prime} =K^{-1}\vec{\ell}^{\,\prime}\cdot\vec{u}^{\,\prime}=K^{-1}/\sqrt{2\;}$.
From this we have  $K^{-1}=\sqrt{2\;}\, \vec{\ell}\cdot 
\vec{u}^{\,\prime}$ and
\begin{equation}\label{Kgammabeta}
  K=\frac{1}{\sqrt{2\;}\,\vec{\ell}\cdot\vec{u}^\prime}
=\frac{1}{\gamma\left(1-{\bf\hat{r}}\cdot{\bs\beta}\right)}\,.
\end{equation}

It should be noticed that we have not completely specified the Lorentz
transformation between the reference frames of observers ${\cal O}$
and ${\cal O}^\prime$; we have specified only the relative velocity,
not the rotations that relate them. From the point of view of the
transformations in Eqs.~\eqref{eq:I1}--\eqref{eq:IV}, we have fixed
the complex parameter $b$ to be zero (since the direction of
$\vec{\ell}$ is fixed) and we have fixed the real parameter $K$ with
Eq.~\eqref{Kgammabeta}.  What remains is the three degrees of freedom
in the parameters $\lambda$ and $a$.  These three degrees of freedom
can be fixed independently for each direction $\theta, \phi$. That is,
$\lambda$ and $a$ can, like $K$ in Eq.~\eqref{Kgammabeta},
be functions of ${\bf\hat{r}}$.  These three degrees of freedom are equivalent
to the freedom to make spatial rotations connecting 
frames ${\cal O} $ and ${\cal O}^\prime$.

The choice of the orientations of these frames is equivalent to a
specification of how the null tetrad legs are chosen. We start by
considering observer ${\cal O}$ with 4-velocity $\vec{u}$ and
radiation in a particular direction. In Eq.~\eqref{tetraddef1} we
chose $\vec{\ell}$ in the direction of propagation, and normalized
$\vec{\ell}$ so that $\vec{u}\cdot\vec{\ell}=1/\sqrt{2\;}$, for
reasons already discussed.  We next, in Eq.~\eqref{tetraddef2},
defined $\vec{n}$ by
\begin{equation}\label{ndef}
  \vec{n}\equiv\sqrt{2\;}\vec{u}-\vec{\ell}\,.
\end{equation}
This choice guarantees that  $\vec{n}$ is null and satisfies the
normalization condition in Eq.~\eqref{eq:normcond}.  It also means
that for any direction of incoming radiation we have a unit
``outward'' spatial vector $\widehat{\bs
  r}\equiv(\vec{n}-\vec{\ell})/\sqrt{2\;}$ that is orthogonal to
$\vec{u}$, and that is parallel to the spatial direction of
propagation. (The vector $\widehat{\bs r}$ is just $\sqrt{2\;}$ times
the projection of $\vec{\ell}$ orthogonal to $\vec{u}$.) In addition,
the form of $\vec{n}$ in Eq.~\eqref{ndef} provides a unit timelike
vector $\widehat{t}\equiv(\vec{n}+\vec{\ell})/\sqrt{2\;}=\vec{u}$ that
is independent of the direction of propagation of the radiation. In
short, we have a tetrad compatible with the $\vec{\ell}$ and $\vec{n}$
legs of Eqs.~\eqref{tetraddef1} and \eqref{tetraddef2}.
We also make the choice 
\begin{equation}
  \label{nprimedef}
 \vec{n}^{\,\prime}
=\sqrt{2\;}\vec{u}^{\,\prime}-\vec{\ell}^{\,\prime} 
\end{equation}
 so that the tetrad for 
${\cal O}^\prime$ is also compatible with the conditions in 
Eqs.~\eqref{tetraddef1} and \eqref{tetraddef2}.

From Eqs.~\eqref{eq:I1}--\eqref{eq:IV}, the
set\cite{orderofXform} of allowed transformations (with $b=0$)
of the tetrad for ${\cal O}^\prime$ in terms of the tetrad for
${\cal O}$ is
\begin{eqnarray}
  \vec{\ell}^{\,\prime}&=& K \vec{\ell}\label{eq:full1}\\
  \vec{m}^{\,\prime}&=& e^{i\lambda} \left(\vec{m}+  a\vec{\ell}\right)\label{eq:full2}\\
  \vec{n}^{\,\prime}&=& K^{-1} \left(\vec{n}+a\vec{m}+a^*\vec{m}^{\,*}
+aa^*K\vec{\ell}\right)\label{eq:full3}\,.
\end{eqnarray}
For simplicity, we first 
 make the restriction that the $z$ axis is chosen in the
direction of $\bs\beta$. (The $\bs\beta$ direction can always be put
in the $z$ direction by using pure rotations.)
We then set equal the
right-hand sides of the expressions for $\vec{n}^{\,\prime}$ given in
Eqs.~\eqref{nprimedef} and \eqref{eq:full3}
\begin{equation}
  \sqrt{2\;}\gamma\{1,{\bs\beta}\}-K\{1,\widehat{r}\}
=K^{-1}\left[
\{1,-\widehat{r}\}
+a\{0,-\widehat{\theta}-i\widehat{\phi}\}
+a^*\{0,-\widehat{\theta}+i\widehat{\phi}\}
+aa^*K\{1,\widehat{r}\}
\right]\,.
\end{equation}
Taking the dot product of this equation, first 
with $\widehat{\theta}$ and then with $\widehat{\phi}$,
for ${\bs\beta}$ in the $z$ direction,
gives us
\begin{eqnarray}
\sqrt{2\;}\gamma\beta\sin\alpha&=&K^{-1}\left[\left(a/\sqrt{2\;}\right)
+\left(a^*/\sqrt{2\;}\right)
\right]  \\
0&=&K^{-1}\left[\left(a/\sqrt{2\;}\right)
\left(i \right)
+\left(a^*/\sqrt{2\;}\right)\left(-i\right)
\right]  \,,
\end{eqnarray}
where $\alpha$ is the angle between the directions of $\bs\beta$ and $\widehat{r}$.
(This is also $\theta$ for $\bs\beta$ in the $z$ direction.)
From this we conclude
\begin{equation}\label{aforzxform}
  a=\beta\gamma K\sin\alpha=\frac{\beta \sin\alpha}{1-\beta\cos\alpha}\,.
\end{equation}
Finally, we note that the direction of $\widehat{\phi}$ is invariant
under a Lorentz boost, since $\widehat{\phi}$ is orthogonal to
$\bs\beta$.  
Since 
$\vec{m}\cdot\widehat{\phi}=\vec{m}^{\prime}\cdot \widehat{\phi}^\prime$,
this means that 
$\vec{m}\cdot\widehat{\phi}=\vec{m}^{\prime}\cdot \widehat{\phi}$.  In
addition, $\vec{m}\cdot\vec{\ell} =0$, so that  the dot product of
$\widehat{\phi}$ with  Eq.~\eqref{eq:full2} 
gives us $\lambda=0$.

In summary, for a pure boost in the $z$ direction the tetrad transformation 
parameters are 
\begin{equation}\label{zxformresults}
  \lambda=b=0\quad\quad\quad K=\frac{1}{\gamma(1-\beta\cos\alpha)}
\quad\quad
a=\frac{\beta\sin\alpha}{(1-\beta\cos\alpha)}\,.
\end{equation}
It is interesting that $\lambda=0$ for this pure boost in the $z$ 
direction. According to
Eq.~\eqref{NPStokesxform} this means that ${\cal S}/{\cal I}$ is
invariant for such a transformation, and therefore
that both observers will agree on the linear polarization of the
radiation.

We now turn to the general transformation, with ${\bs\beta}$ not necessarily in 
the $z$ direction.  We have already constructed $\vec{\ell}^{\,\prime}$ and $\vec{n}^{\,\prime}$
from the propagation vector and the 4-velocity $\vec{u}^{\,\prime}$. What remains 
is to find $\vec{m}^{\,\prime}$. 
To do this we note that from the forms of $\vec{m}$ and $\vec{\ell}$ given
in Sec.~\ref{sec:NPform} we have
\begin{equation}\label{mfromell}
  \vec{m}\left(\theta,\phi\right)
=-\,\left(\partial_\theta+(i/\sin\theta)\partial_\phi\right)
\vec{\ell}\left(\theta,\phi\right)\,.
\end{equation}
An analogous relationship must obtain in the primed frame, so we can write
\begin{equation}
  \vec{m}^{\prime}\left(\theta^{\prime},\phi^{\prime}\right)
=-\,\left(\partial_{\theta^{\prime}}+(i/\sin\theta^{\prime})\partial_{\phi^{\prime}}\right)
\vec{\ell}^{\,\prime}\left(\theta^{\prime},\phi^{\prime}\right)
=-\,\left(\partial_{\theta^{\prime}}+(i/\sin\theta^{\prime})\partial_{\phi^{\prime}}\right)
K(\theta,\phi)\vec{\ell}\left(\theta,\phi\right)\,.
\end{equation}
In order to compute the right hand side of this equation, we must know 
$\theta,\phi$ as functions of 
$\theta^{\prime},\phi^{\prime}$ for an arbitrary Lorentz transformation.

The
general Lorentz transformation, for an arbitrary boost combined with
an arbitrary rotation, requires the specification of six parameters.
It is convenient here to utilize the description and parameterization
related to the ``aberration transform,'' as described e.g.\,, in
Ref.~\cite{NPpedapaper} with a slightly different
notation\cite{caps}. In this description the direction of photon
propagation $\theta^\prime,\phi^\prime$ in frame ${\cal O^\prime}$ is
related to the direction $\theta,\phi$ for the same photon in frame
${\cal O}$ by
\begin{equation}\label{aberration}
  e^{i\phi^\prime}\cot{(\theta^\prime/2)}
=\frac{
A\, e^{i\phi}\cot(\theta/2)+B
}{
C\,e^{i\phi}\cot(\theta/2)+D
}\,.
\end{equation}
Here $A,B,C,D$ are complex numbers.  The eight degrees of freedom in
these complex numbers are constrained by the two conditions in
$AD-BC=1$, leaving six unconstrained degrees of freedom
in the transformation \eqref{aberration}. Any proper (non-time
reversing, non-parity reversing) Lorentz transformation can be
represented with these six degrees of freedom.

Carrying out the differentiations indicated gives 
\begin{equation}\label{genmxform}
  \vec{m}^{\prime}\left(\theta^\prime,\phi^\prime\right)=e^{i(\sigma-\phi+\phi^\prime)}
\left(\vec{m}\left(\theta,\phi\right)+ J(\theta,\phi)\,\vec{\ell}\left(\theta,\phi\right)\right)\,,
\end{equation}
where 
\begin{equation}\label{edth3}
  J(\theta,\phi)=-\,\left(\partial_\theta+(i/\sin\theta)\partial_\phi\right)
\left[
\log{K\left(\theta,\phi\right)}
\right]\,,
\end{equation}
and
\begin{equation}\label{sigma}
  e^{i\sigma}=\sqrt{
\frac{{C}^*e^{-i\phi}\cot(\theta/2)+{D}^*}{C\;e^{i\phi}\cot(\theta/2)\,+{D}}
}\,.
\end{equation}
The values of $a$ and $\lambda$ are inferred from a comparison of the 
result in Eq.~\eqref{genmxform} and Eq.~\eqref{eq:full2}.

The special case of the pure boost in the $z$ direction is given by $B=C=0$
and $D=1/A$, with $A$ a real number. Equation~\eqref{sigma} gives $\sigma=0$. Since
$K$, from Eq.~\eqref{zxformresults}, in a slighly different notation, is
$K=[\gamma(1-\beta\cos\theta)]^{-1}$ we have, from Eq.~\eqref{edth3},
\begin{equation}
  J=\partial_\theta[\log{\gamma(1-\beta\cos\theta)}]
=\frac{\beta\sin\theta}{1-\beta\cos\theta}\,,
\end{equation}
which means, according to Eq.~\eqref{genmxform} that 
\begin{equation}
  \vec{m}^\prime=  \vec{m}+\frac{\beta\sin\theta}{1-\beta\cos\theta}\vec{\ell}\
\end{equation}
in agreement with Eqs.~\eqref{eq:full2},\eqref{zxformresults} for the 
case of a pure boost in the $z$ direction.

\section{Representations of the Lorentz group and the relevance to the
  CMB}\label{app:GGV}

Following the beautiful theory of representations of the Lorentz group
developed by Gel'fand, Graev and Velinkin (GGV)\cite{GGV}, one finds
that the different Stokes fields lying on the celestial sphere, lie in
different vector spaces of infinite dimensional representations of the
Lorentz group. In the notation of GGV, representations are given by
homogeneous functions of two complex variables
($z_{1,}z_{2,}\overline{z}_{1,}\overline{z}_{2,}$), each
representation being labeled by the homogeneity degree,
($n_{1}-1,n_{2}-1$) of each pair of the variables. We consider only
the so-called integer representations, for which $n_{1}$ and $n_{2}$
are either both positive or both negative integers. These
representations, via homogeneous functions, can be mapped into
functions on the sphere\ with well defined spin and conformal weights,
$s$ and $w$. The $s$ and $w$ are related to $n_{1}$ and $n_{2}$
by\cite{HeldNewmanPosadas70}
\begin{equation}
(n_{1},n_{2})=(w-s+1,w+s+1).  \label{A1}
\end{equation}

A variety of results can be extracted from the GGV work. If $n_{1}$ and $%
n_{2} $ are either both positive or both negative, one can find (or
construct), from the infinite dimensional representation spaces, specific
finite dimensional vector spaces that are equivalent to the standard tensor
representations. (A simple example is that the $\ell=0$ harmonic coefficient of
a $w=-2,s=0$ function is a Lorentz scalar. See below.)

We review the basic argument and results for negative integer
representations, taking for simplicity the special case of $s=0$.
First, we note that the area element on the unit sphere (or equivalently, the
solid angle as seen on the celestial sphere)\cite{HeldNewmanPosadas70}
\begin{equation}
d\Omega =\sin \theta d\theta d\varphi ,  \label{A2}
\end{equation}%
transforms (under the aberration transformation of Eq.~\eqref{aberration}) as
\begin{equation}
d\Omega ^{\prime }=K^{2}d\Omega .  \label{A3}
\end{equation}%
Second, we consider functions $F_{(-n-2)}$ and $G_{(n)}$ on the
sphere, respectively with boost weight $-n-2$ and $n$, that is with
Lorentz transformation properties
\begin{equation}
F_{(-n-2)}^{\,\prime } =K^{-n-2}F_{(-n-2)}\quad\quad
G_{(n)}^{\,\prime } =K^{n}G_{(n)}\,.  
\end{equation}
The integral 
\begin{equation}
  \int F_{(-n-2)}G_{(n))}\,d\Omega
\end{equation}
over the 2-sphere is therefore a Lorentz invariant:
\begin{equation}
  \int F_{(-n-2)}G_{(n))}\,d\Omega=  \int F^\prime_{(-n-2)}G_{(n))}^\prime\,d\Omega^\prime\,.
\end{equation}
In a similar manner, we can create Lorentz vectors and tensors.

The following are several simple specific examples:

\medskip
(a) If we choose $G_{(0)}=1$ and choose $F_{(-2)}$ arbitrarily, we have 
 that
\begin{equation}
{\rm Harmonic}_{(\ell=0)}=\int F_{(-2)}d\Omega 
=\int F_{(-2)}^{\,\prime }d\Omega^{\prime }  \,,   \label{A7}
\end{equation}%
is a Lorentz invariant. That is, we have that 
 that the monopole, the $\ell=0$ harmonic 
coefficient, of $F_{(-2)}$\ is Lorentz invariant.

\medskip
(b) If we choose $G_{(1)}=\ell^{a}$ and choose $F_{(-3)}$
arbitrarily we have that
\begin{equation}
w^{a}\equiv \int F_{(-3)}l^{a}d\Omega\,,  \label{A8}
\end{equation}
is a Lorentzian 4-vector  extracted from $F_{(-3)}$.

\medskip
(c) If we choose $G_{(2)}=\ell^{a}\ell^{b}$ and choose $F_{(-4)}$ arbitrarily we
have that
\begin{equation}
w^{ab}\equiv \int F_{(-4)}\ell^{a}\ell^{b}d\Omega\,,  \label{A9}
\end{equation}%
a trace-free symmetric tensor extracted from $F_{(-4)}.$

A subtle issue arises in the above constructions. In the body of the paper we took
the components $\ell^{a}$ in one particular Minkowski coordinate system to have  the
canonical form
\begin{equation}
\ell^{a}=\frac{\sqrt{2}}{2}(1,\sin \theta \cos \varphi ,\sin \theta \sin
\varphi ,\cos \theta )\,.  \label{canonical}
\end{equation}%
For $\vec{\ell}$ in the primed frame, we must have $\vec{\ell}^{\;\prime}=K{\vec\ell}$, 
and hence the components of 
$\vec{\ell}^{\,\prime}$ must be ($\theta,\phi$-dependent) linear combinations 
of the same harmonics, i.e., of $(1,\sin \theta\cos \varphi,
\sin \theta\sin \varphi,\cos \theta)$\,.
But the components of $\vec{\ell}^{\;\prime}$, must have the canonical form 
$2^{-1/2}(1,\sin \theta ^{\prime }\cos \varphi ^{\prime
},\sin \theta ^{\prime }\sin \varphi ^{\prime },\cos \theta ^{\prime })$\,,
in the primed Minkowski frame. This means that a Lorentz transformation of coordinates
\begin{equation}
  \ell^{\prime a^{\prime }}=L_{a}^{a^{\prime }}\ell^{\prime a}=\frac{\sqrt{2}}{2}%
(1,\sin \theta ^{\prime }\cos \varphi ^{\prime },\sin \theta ^{\prime }\sin
\varphi ^{\prime },\cos \theta ^{\prime })
\end{equation}
is necessary to give the canonical form for the components $\vec{\ell}^{\,\prime}$.
This in turn implies, via $\ell_{a^{\prime }}^{\prime }w^{\prime a^{\prime
}}=\ell_{a}w^{a},$ a (coordinate) Lorentz transformation on the $w^{\prime a},$ 
\[
w^{\prime a^{\prime }}=L_{a}^{a^{\prime }}w^{\prime a}.
\]%
\qquad \qquad \qquad 

These relations are easily translated to the Stokes fields,
$\mathcal{I}%
\mathcal{(\theta }$,$\mathcal{\varphi )}$,
$\mathcal{S}\mathcal{(\theta }$,
$\mathcal{\varphi )}$ and $\mathcal{V}\mathcal{(\theta
}$,$\mathcal{\varphi )} $ discussed in the text and, as well, to the
closely associated quantity,   the intensity per unit
solid angle per unit frequency interval $\mathcal{I}_{\nu
}\mathcal{(\theta },\mathcal{\varphi )}.$ The 
$\mathcal{I}_{\nu}$ transforms as with $s=0$,
amd $w=-3$\,.

Since $\mathcal{I(\theta }$,$\mathcal{\varphi )}$ has boost weight
$w=-2$ and spin weight $s=0,$ a formal set of relations arise from the
GGV work, and a set of Lorentzian tensors can be directly extracted as follows
from $\mathcal{I}$,  $\mathcal{I}_{\nu }$, or more generally
from $\mathcal{I}/\nu^N$\,.

(1) From Eq.~\eqref{A7}, it follows that the coefficient of the
$\ell=0$ harmonic, the monopole of $\mathcal{I}$, is a Lorentz
invariant, $T_{0}$.

(2) From Eq.~\eqref{A8}, in the spherical harmonic expansion of the
$w=-3$ quantity $\mathcal{I}^{\frac{3}{2}},$  the
coefficients of the ($\ell=0,1)$ harmonics form and transform as a
Lorentzian four-vector, say $T^{a}$.
In the same manner, the coefficients of the ($\ell=0,1)$ harmonics in
$\mathcal{I}_{\nu}$ determine a four-vector, $v^{a}$ that can be
considered as defining the rest-frame for the radiation bath.

(3) From Eq.~\eqref{A9}, the spherical harmonic expansion of 
either of the $w=-4$
quantities
 $\mathcal{I}^{2}$, $\nu\,\mathcal{I}$,  the coefficients of the 
($\ell=0,1,2)$ harmonics form and transform as a Lorentzian trace-free, symmetric
two-index tensor, $T^{ab}$ and $V^{ab}$ respectively.

(4) These results generalize to all higher powers of $\mathcal{I}$, i.e. to
all cases of integer $w\,<-4,$ and $s=0.$

\smallskip These results allow us to assign a physical meaning for the
vector defined in (2). The quantity $\mathcal{I}_{\nu}$, \textit{the
  intensity per unit solid angle per unit frequency interval, }is the
physical quantity that is usually measured in CMB observations. Since
it is a quantity of weight $%
w=-3\ $\ it has, as was mentioned above, an invariantly defined
Lorentzian four-vector, $v^{a}$, the $\ell=0,1$ harmonics. Since the
monopole is far larger than the dipole term, this vector is time-like
and can be identified with the observers velocity with respect to the
CMB background rest-frame.  To use the Lorentzian vector $v^{a}\ $as a
`rest-frame' for the background radiation field one finds a moving
frame in which $\mathcal{I}_{\nu}$ has no dipole component.
More explicitly, the velocity of the rest frame, thus defined, is
\begin{equation*}
v^{a}\equiv \int \mathcal{I}_{\nu}l^{a}d\Omega 
\end{equation*}%
This in turn could be used to find the $K$ and then determine the rest-frame
distribution%
\begin{equation*}
\mathcal{I}_{\nu }^{\prime }=K^{-3}\mathcal{I}_{\nu}.
\end{equation*}


\bibliographystyle{aps}

\begin{thebibliography}{20}
\bibitem[{Fmu()}]{Fmunuconvention}
\bibinfo{note}{The overall signs depend on conventions, but the overall signs
  will not matter in the considerations that follow.}
\bibitem[{orp()}]{orplus2}
\bibinfo{note}{We can, alternatively, replace ${\cal
  S}\equiv2\left\langle\Phi_2\Phi_2\right\rangle$ by ${\cal
  S}^*\equiv\left\langle\Phi_2^*\Phi_2^*\right\rangle=Q+iU$ which is to be
  analyzed in spin-weight +2 harmonics.}
\bibitem[{ord()}]{orderofXform}
\bibinfo{note}{The equations here apply if the Class II transformations are
  made first, followed by the transformations in Class III and IV.}
\bibitem[{cap()}]{caps}
\bibinfo{note}{Here we use capital $A,B,C,D$ rather than the more common lower
  case $a,b,c,d$ used in Ref.~\cite{NPpedapaper} to avoid confusion with the
  parameter $a$ introduced in Eqs.~\eqref{firstclass2} and
  \eqref{firstclass4}.}
\end{thebibliography}


\end{document}